\documentclass{elsart}

\usepackage{cite}
\usepackage{graphicx}
\usepackage{dcolumn}

\begin{document}

\begin{frontmatter}

\title{Three PT-symmetric Hamiltonians with completely different spectra}

\author{Francisco M. Fern\'{a}ndez\thanksref{FMF}} \and \author{Javier Garcia}

\address{INIFTA (UNLP, CCT La Plata-CONICET), Divisi\'on Qu\'imica Te\'orica,
Blvd. 113 S/N,  Sucursal 4, Casilla de Correo 16, 1900 La Plata,
Argentina}

\thanks[FMF]{e--mail: fernande@quimica.unlp.edu.ar}

\begin{abstract}
We discuss three Hamiltonians, each with a central-field part
$H_{0}$ and a PT-symmetric perturbation $igz$. When $H_{0}$ is the
isotropic Harmonic oscillator the spectrum is real for all $g$
because $H$ is isospectral to $H_{0}+g^{2}/2$. When $H_{0}$ is the
Hydrogen atom then infinitely many eigenvalues are complex for all
$g$. If the potential in $H_{0}$ is linear in the radial variable
$r$ then the spectrum of $H$ exhibits real eigenvalues for
$0<g<g_{c}$ and a PT phase transition at $g_c$.
\end{abstract}


\end{frontmatter}

\section{Introduction}

\label{sec:intro}

It is known since long ago that some non-Hermitian operators may exhibit
real eigenvalues\cite{CGM80,A95}. This fact remained a somewhat exotic
mathematical subject till Bender and Boettcher\cite{BB98} suggested that
those operators may exhibit unbroken parity-time (PT) symmetry. From then on
the problem quickly developed into a prolific field of research\cite{B07}
(and references therein).

In a roughly general way we may say that most of the studied problems are
represented by Hamiltonian operators of the form $H=H_{0}+\lambda H^{\prime
} $, where $H_{0}$ is parity-invariant $PH_{0}P=H_{0}$ and $H^{\prime }$ is
parity antisymmetric $PH^{\prime }P=-H^{\prime }$, where $P$ is the parity
operator. If $\lambda =ig$ is imaginary (where $g$ is obviously real) then $%
H $ is PT symmetric: $PTHPT=H$, where $T$ is the time-reversal operator\cite
{P65}.

In the beginning, most of the models studied were mainly one-dimensional\cite
{BB98,B07,FGZ98,FGRZ99} in which case $H_{0}$ only exhibits parity symmetry
and its eigenfunctions $\psi ^{(0)}$ are even or odd: $P\psi ^{(0)}=\pm \psi
^{(0)}$ but later the researchers began to look for multidimensional examples%
\cite{BDMS01, NA02, N02, N05, BTZ06, W09, BW12, HV13}. It was suggested that
space-time (ST) symmetry could be a suitable generalization of the PT one%
\cite{KC08}. In this case $SH_{0}S=H_{0}$ and $SH^{\prime }S=-H^{\prime }$,
where $S$ is a unitary operator such that $S^{\dagger }=S^{-1}=S$. Most of
the effort was devoted to find new multidimensional non-Hermitian
Hamiltonians with real spectra.

In the multidimensional case $H_{0}$ and $H$ may exhibit more complex
symmetry that is conveniently described by means of group theory. In this
way Fern\'{a}ndez and Garcia\cite{FG14a,FG14b} and Amore et al\cite
{AFG14b,AFG14c} found that some ST-symmetric Hamiltonians exhibit broken ST
symmetry for all values of $g$. The main conjecture was that ST symmetry may
be unbroken for some values of $g$ provided that $S$ is the only member of a
class in the point group for $H_{0}$\cite{AFG14b}. This appeared to be the
case when $S=P$.

The purpose of this paper is the discussion of three PT-symmetric
Hamiltonians for which $H_{0}=p^{2}/2+V(r)$ exhibits central-field
symmetry and $H^{\prime }=z$. The resulting Hamiltonian $H$
exhibits cylindrical symmetry and may be viewed as a kind of Stark
effect with imaginary electric field. In
Section~\ref{sec:PTsymmetry} we outline the main ideas of PT
symmetry as well as a simple argument based on perturbation
theory\cite {FG14a,FG14b,AFG14b,AFG14c}. In
Section~\ref{sec:Stark_gen} we briefly discuss the general case.
In sections \ref{sec:HO}, \ref{sec:HA} and \ref
{sec:linear} we show that the models with $V(r)=r^{2}/2$, $V(r)=-1/r$, and $%
V(r)=r$, respectively, exhibit completely different spectra.
Finally, in Section~\ref{sec:conclusions} we summarize the main
results of the paper and draw conclusions.

\section{Parity-time symmetry}

\label{sec:PTsymmetry}

Let $A=PT=A^{-1}$ be the antiunitary operator given by the product of the
parity $P$ and time-reversal $T$ operators\cite{P65,BB98}. The Hamiltonian
operator $H$ is said to be $PT$ symmetric if
\begin{equation}
AHA^{-1}=H.  \label{eq:AHA=H}
\end{equation}

If
\begin{equation}
H\psi =E\psi ,  \label{eq:Hpsi=Epsi}
\end{equation}
then
\begin{equation}
AH\psi =AHA^{-1}A\psi =HA\psi =AE\psi =E^{*}A\psi .  \label{eq:AHpsi}
\end{equation}
If $A\psi =a\psi $, $a$ being a complex number, then we say that PT symmetry
is unbroken and $E=E^{*}$. It is not difficult to prove that $|a|=1$.
Fern\'{a}ndez and Garcia\cite{FG14c} found a case in which $A\psi \neq a\psi
$ and still $E$ is real. They proposed the supposedly more general condition
$HA\psi =EA\psi $ for the occurrence of real spectrum; that is to say, when $%
\psi $ and $A\psi $ are two linearly independent eigenfunctions of $H$ with
the same eigenvalue $E$. This situation does not take place unless the
spectrum of $H$ is degenerate. However, further analysis reveals that both
conditions are equivalent. In fact, if we choose $\varphi =c_{1}\psi
+c_{2}A\psi $, where $c_{2}^{*}=ac_{1} $ and $c_{1}^{*}=ac_{2}$, then $%
A\varphi =a\varphi $. It is worth adding that none of these conditions is of
practical utility to predict whether $H$ will have real eigenvalues or not
because one commonly ignores the effect of $A$ on the eigenvectors of $H$.

Most of the examples studied so far are of the form
\begin{equation}
\;H=H_{0}+\lambda H^{\prime },  \label{eq:H_general}
\end{equation}
where
\begin{equation}
PH_{0}P=H_{0},\;TH_{0}T=H_{0},\;PH^{\prime }P=-H^{\prime },\;TH^{\prime
}T=H^{\prime }  \label{eq:PH0P=H0_etc}
\end{equation}
and $\lambda =ig$, where $g$ is real. Since $T\lambda H^{\prime }T=\lambda
^{*}TH^{\prime }T=-\lambda H^{\prime }$ then $AHA=H$. Some useful
information on the spectrum of $H$ is given by the perturbation series
\begin{equation}
E=\sum_{j=0}E^{(j)}\lambda ^{j},  \label{eq:E_series}
\end{equation}
because if at least one coefficient of odd order $E^{(2i+1)}$ is nonzero
then $E$ is expected to be complex for sufficiently small $g$. In such a
case the PT-phase transition\cite{BW12} takes place at the trivial Hermitian
limit $g=0$. If we write $H(\lambda )\psi _{m}=E_{m}(\lambda )\psi _{m}$
then $PH(\lambda )\psi _{m}=PH(\lambda )PP\psi _{m}=H(-\lambda )P\psi
_{m}=E_{m}(\lambda )P\psi _{m}$. If $\psi _{m}$ and $P\psi _{m}$ are
linearly dependent, then $E_{m}(-\lambda )=E_{m}(\lambda )$ and all the
perturbation corrections of odd order vanish; otherwise $P\psi _{m}=\psi
_{n} $, $E_{m}(-\lambda )=E_{n}(\lambda )$ and we cannot draw a conclusion
so easily. The latter case may only take place when the spectrum of $H$ is
degenerate. In many cases it suffices to calculate the simplest,
straightforward perturbation correction of first order $E^{(1)}$\cite
{FG14a,FG14b,AFG14b,AFG14c}.

\section{Stark effect}

\label{sec:Stark_gen}

Consider the Hamiltonian operator
\begin{equation}
H=-\frac{1}{2}\nabla ^{2}+V(r)+\lambda z,  \label{eq:H=cent_field+H'}
\end{equation}
where $V(r)$ is spherically symmetric (depends only on $r$). The
eigenfunctions of $H_{0}=H(\lambda =0)$
\begin{equation}
H_{0}\psi _{\nu \,l\,m}^{(0)}=E_{\nu \,l\,}^{(0)}\psi _{\nu \,l\,m}^{(0)},
\label{eq:H0psi0}
\end{equation}
are also eigenfunctions of the angular momentum operators $L^{2}$ and $L_{z}$%
\begin{eqnarray}
L^{2}\psi _{\nu \,l\,m}^{(0)} &=&l(l+1)\psi _{\nu \,l\,m}^{(0)},  \nonumber
\\
L_{z}\psi _{\nu \,l\,m}^{(0)} &=&m\psi _{\nu \,l\,m}^{(0)},  \nonumber \\
l &=&0,1,\ldots ,\;m=0,\pm 1,\ldots ,\pm l.  \label{eq:Lpsi0}
\end{eqnarray}
In spherical coordinates the eigenfunctions can be factored as
\begin{equation}
\psi _{\nu \,l\,m}^{(0)}(r,\theta ,\phi )=R_{\nu \,l}(r)Y_{l}^{m}(\theta
,\phi ),  \label{eq:psi0=RY}
\end{equation}
where $R_{\nu \,l}(r)$ is the radial part, $\nu =0,1,\ldots $, is the radial
quantum number and $Y_{l}^{m}(\theta ,\phi )$ are the spherical harmonics.
Since the eigenvalues of $H_{0}$ do not depend on $m$ they are at least $%
(2l+1)$-fold degenerate.

The perturbation $H^{\prime }=z$ breaks the degeneracy of the spectrum of $%
H_{0}$ but the states with $m>0$ remain two-fold degenerate because the
eigenvalues of $H$ do not depend on the sign of the magnetic quantum number $%
m $.

Since
\begin{equation}
P\psi _{\nu \,l\,m}^{(0)}=(-1)^{l}\psi _{\nu \,l\,m}^{(0)},  \label{eq:Ppsi0}
\end{equation}
and $PzP=-z$ the matrix elements
\begin{equation}
z_{\nu \,l\,m}^{\nu ^{\prime }l^{\prime }m}=\left\langle \psi _{\nu
\,l\,m}^{(0)}\right| z\left| \psi _{\nu ^{\prime }l^{\prime
}m}^{(0)}\right\rangle ,  \label{eq:z_mat_el}
\end{equation}
are zero when $l-l^{\prime }$ is even. The perturbation corrections of first
order to the energy $E_{\nu lm}^{(1)}$ are given by the eigenvalues of the
matrix with elements $z_{\nu \,l\,m}^{\nu ^{\prime }l^{\prime }m}$. We will
discuss three examples in the subsequent sections.

\section{Isotropic harmonic oscillator}

\label{sec:HO}

When

\begin{equation}
V(r)=\frac{1}{2}r^{2}  \label{eq:V(r)=r^2/2}
\end{equation}
the Schr\"{o}dinger equation is exactly solvable and the eigenfunctions and
eigenvalues are given by
\begin{eqnarray}
\psi _{n_{1}\,n_{2}\,n_{3}}(x,y,z) &=&\varphi _{n_{1}}(x)\varphi
_{n_{2}}(y)\varphi _{n_{3}}(z+\lambda ),  \nonumber \\
E_{k} &=&\left( k+\frac{3}{2}\right) -\frac{1}{2}\lambda
^{2},\;k=n_{1}+n_{2}+n_{3},  \nonumber \\
n_{1},n_{2},n_{3} &=&0,1,\ldots ,  \label{eq:IHO_eigenv_eigenf}
\end{eqnarray}
where $\varphi _{n}(q)$ is an eigenfunction of the one-dimensional harmonic
oscillator $H_{HO}=-\frac{1}{2}\frac{d^{2}}{dq^{2}}+\frac{1}{2}q^{2}$.

Since
\begin{eqnarray}
A\psi _{n_{1}\,n_{2}\,n_{3}}(x,y,z) &=&\psi
_{n_{1}\,n_{2}\,n_{3}}(-x,-y,-z)^{*}=\varphi _{n_{1}}(-x)\varphi
_{n_{2}}(-y)\varphi _{n_{3}}(-z+\lambda ^{*})  \nonumber \\
&=&(-1)^{k}\psi _{n_{1}\,n_{2}\,n_{3}}(x,y,z),  \label{eq:Apsi_HO}
\end{eqnarray}
then the PT symmetry is unbroken for all $g$ which accounts for the fact
that the eigenvalues in equation (\ref{eq:IHO_eigenv_eigenf}) are real for
all $g$.

Although in this case the approximate analysis based on perturbation theory
may appear to be unnecessary we carry it out anyway merely for comparison
purposes. To begin with, note that $P\psi
_{n_{1}\,n_{2}\,n_{3}}^{(0)}(x,y,z)=(-1)^{k}\psi
_{n_{1}\,n_{2}\,n_{3}}^{(0)}(x,y,z)$. The perturbation correction of first
order to a given energy level $E_{k}^{(0)}$ is given by matrix elements of
the form
\begin{equation}
z_{n_{1}\,n_{2}\,n_{3}}^{m_{1}\,m_{2}\,m_{3}}=\left\langle \psi
_{n_{1}\,n_{2}\,n_{3}}^{(0)}\right| z\left| \psi
_{m_{1}\,m_{2}\,m_{3}}^{(0)}\right\rangle ,
\end{equation}
that vanish for all degenerate states because $%
k=n_{1}+n_{2}+n_{3}=m_{1}+m_{2}+m_{3}$. Therefore, $E_{k}^{(1)}=0$ for all
the states of the PT Stark effect in the isotropic harmonic oscillator. This
result is consistent with the form of the exact eigenvalues (\ref
{eq:IHO_eigenv_eigenf}) that depend on $g^{2}$.

There is another way to prove that the PT symmetry for this problem remains
unbroken for all values of $g$. The proof is based on the fact that $H$ can
be written in terms of a similarity transformation of $H_{0}$:
\begin{equation}
H=UH_{0}U^{-1}+\frac{g^{2}}{2},\;U=e^{-gp_{z}},  \label{eq:similarity}
\end{equation}
where $p_{z}=-i\frac{d}{dz}$. Obviously, $H_{0}$ and $UH_{0}U^{-1}$ are
isospectral\cite{F15}.

\section{Hydrogen atom}

\label{sec:HA}

The unperturbed eigenvalues for the Coulomb interaction
\begin{equation}
V(r)=-\frac{1}{r},  \label{eq:V(r)=-1/r}
\end{equation}
are given by
\begin{equation}
E_{n}^{(0)}=-\frac{1}{2n^{2}},\;n=\nu +l+1.
\end{equation}
Therefore, there are pairs of degenerate states $\psi _{\nu lm}^{(0)}$, $%
\psi _{\nu ^{\prime }l^{\prime }m}^{(0)}$ for which $l-l^{\prime }=\nu
^{\prime }-\nu $ is odd and the corresponding matrix elements $z_{\nu
\,l\,m}^{\nu ^{\prime }l^{\prime }m}$ (\ref{eq:z_mat_el}) are nonzero. In
such cases, which for real $\lambda $ give rise to what is commonly known as
linear Stark effect\cite{BS57, LL65}, the perturbation correction of first
order is nonzero and the eigenvalues of $H$ are complex for $g\neq 0$.

The Schr\"{o}dinger equation for this problem is separable in parabolic
coordinates and the exact calculation of the perturbation corrections in
terms of the parabolic quantum numbers $n_{1}=0,1,\ldots $, $%
n_{2}=0,1,\ldots $ and $m=0,\pm 1,\ldots $ is
straightforward\cite{F00}. It is customary to write the
perturbation series
\begin{equation}
E_{nq|m|}=\sum_{j=0}^{\infty }E_{nq|m|}^{(j)}\lambda ^{j},
\end{equation}
in terms of the quantum numbers $n=n_{1}+n_{2}+|m|+1$ and $q=n_{1}-n_{2}$%
\cite{F00}. All the coefficients of odd order vanish when $q=0$ but the
states with $q\neq 0$ are expected to be complex when $g\neq 0$.

\begin{figure}[H]
\begin{center}
\includegraphics[width=6cm]{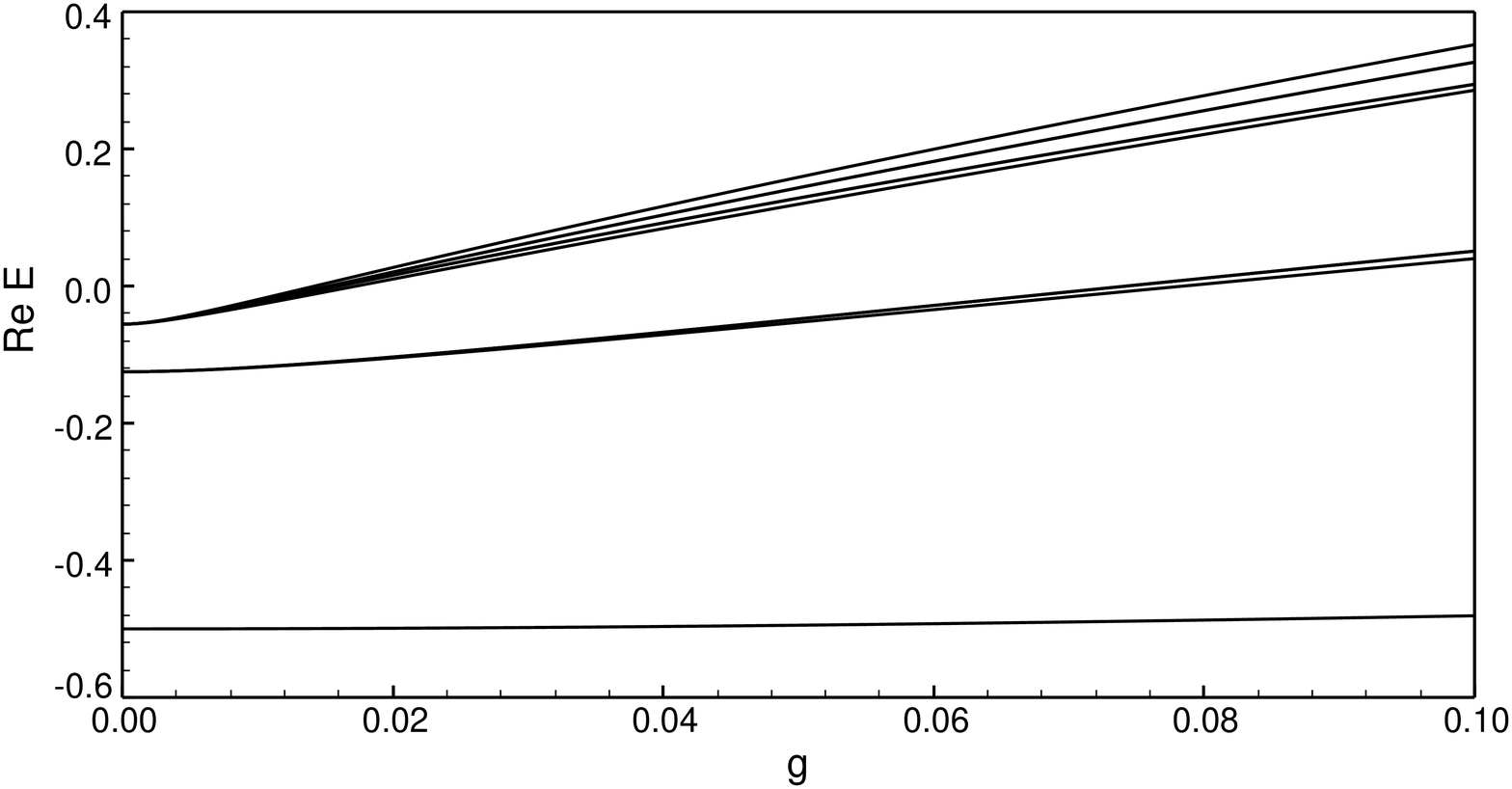} %
\includegraphics[width=6cm]{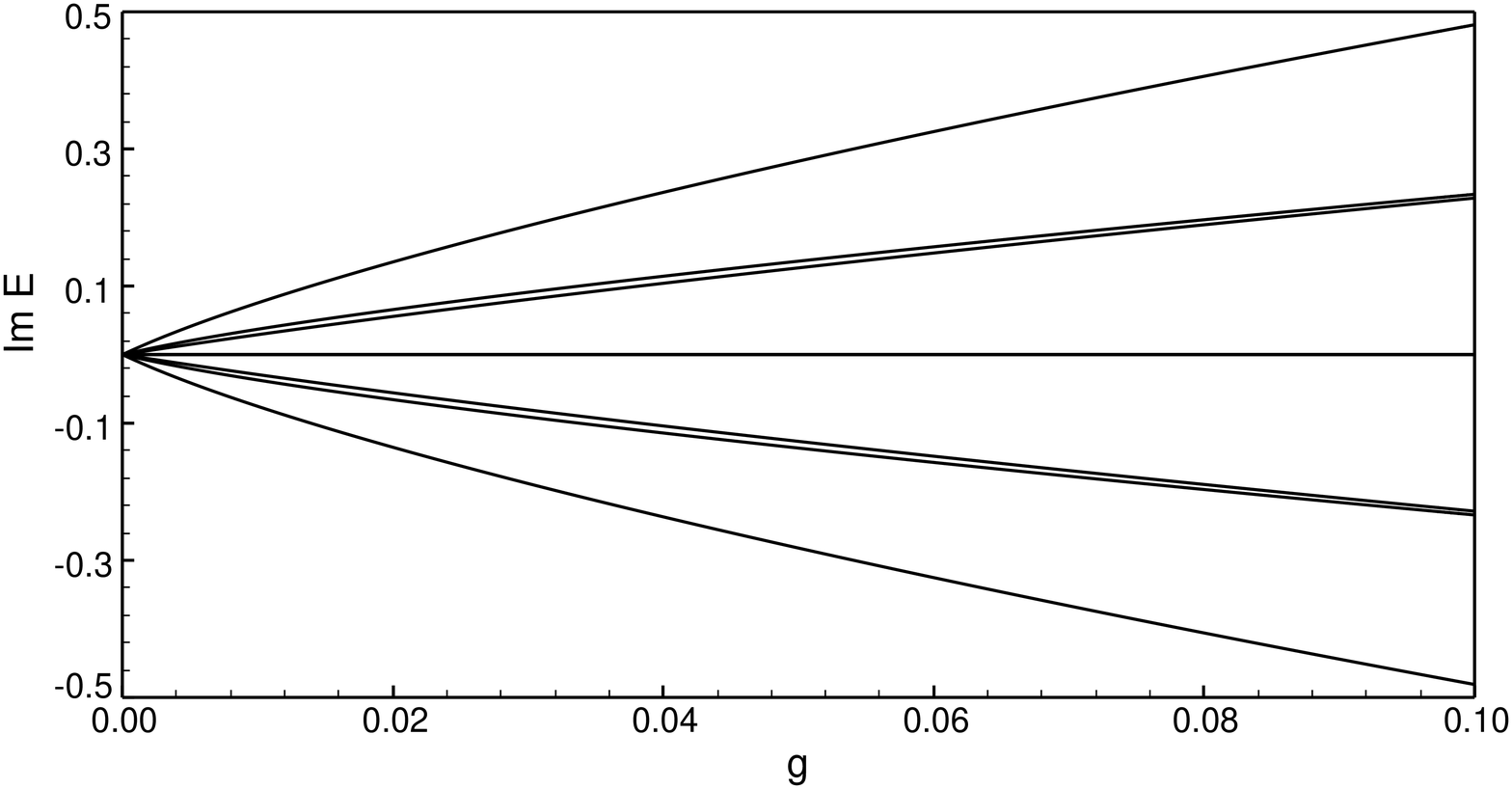}
\end{center}
\caption{Real and imaginary parts of the lowest eigenvalues of the
PT-symmetric Stark effect in hydrogen}
\label{fig:Stark}
\end{figure}

The argument based on perturbation theory just outlined is sufficient to
conclude that this model exhibits complex eigenvalues when $g\neq 0$ and
that the PT phase transition\cite{BW12} takes place at the trivial Hermitian
limit $g=0$. Nevertheless, we will show some numerical results to illustrate
the point. Here we choose the most efficient method of Benassi and Grecchi%
\cite{BG80} that is based on the separation of the Schr\"{o}dinger equation
in squared-parabolic coordinates. Since the details of this approach have
been given elsewhere\cite{BG80,FG15a,FG15b}, here we just show the results.
Figure~\ref{fig:Stark} shows the real and imaginary parts of the lowest
eigenvalues. It is clear that the PT phase transition takes place at the
trivial Hermitian limit as already argued above.

The remarkable difference between the spectra of this problem and the
previous one can be traced back to the symmetry of $H_{0}$. The general
central-field model is invariant under the group O(3) while, on the other
hand, the hydrogen atom is invariant under the group O(4)\cite{L68}. Such
higher symmetry is due to the conservation of the Runge-Lenz vector in the
latter model. Thus, the higher symmetry of $H_{0}$ appears to be the reason
why the PT symmetry is broken for all $g$ in the perturbed hydrogen atom.
While the $k$-th harmonic-oscillator eigenvalue $E_{k}^{(0)}$ is $\frac{%
(k+1)(k+2)}{2}$-fold degenerate, the $n$-th eigenvalue of the hydrogen atom $%
E_{n}^{(0)}$ is $n^{2}$-fold degenerate. The greater degeneracy of the
latter model allows the appearance of nonzero matrix elements $z_{\nu
\,l\,m}^{\nu ^{\prime }l^{\prime }m}$ and nonzero perturbation corrections
of first order.

\section{Linear potential}

\label{sec:linear}

As a nontrivial example we consider the linear potential
\begin{equation}
V(r)=r.  \label{eq:V(r)=r}
\end{equation}
In this case we cannot solve the eigenvalue equation for $H_{0}$ exactly but
we can nevertheless calculate the perturbation correction of first order to
any energy level $E_{\nu \,l}^{(0)}$ because it is determined by matrix
elements of the form
\begin{equation}
z_{\nu \,l\,m}^{\nu \,l\,m^{\prime }}=\left\langle \psi _{\nu
\,l\,m}^{(0)}\right| z\left| \psi _{\nu \,l\,m^{\prime }}^{(0)}\right\rangle
,
\end{equation}
which vanish for all sets of quantum numbers as argued in
Section~\ref {sec:Stark_gen}. Therefore, $E_{\nu
\,l\,|m|}^{(1)}=0$ and there is a chance that PT symmetry may be
unbroken for sufficiently small $g$.

\begin{figure}[H]
\begin{center}
\includegraphics[width=6cm]{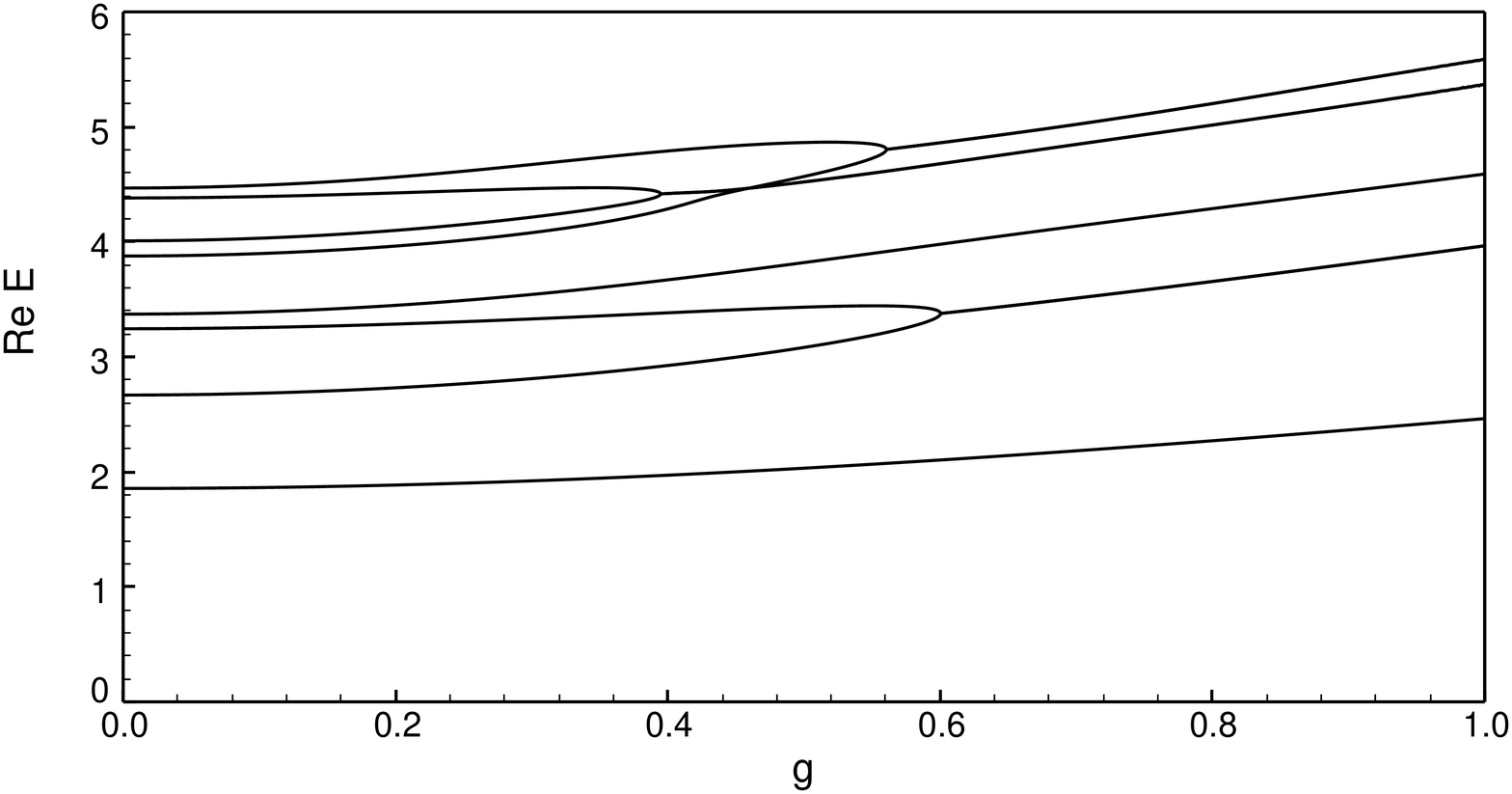} %
\includegraphics[width=6cm]{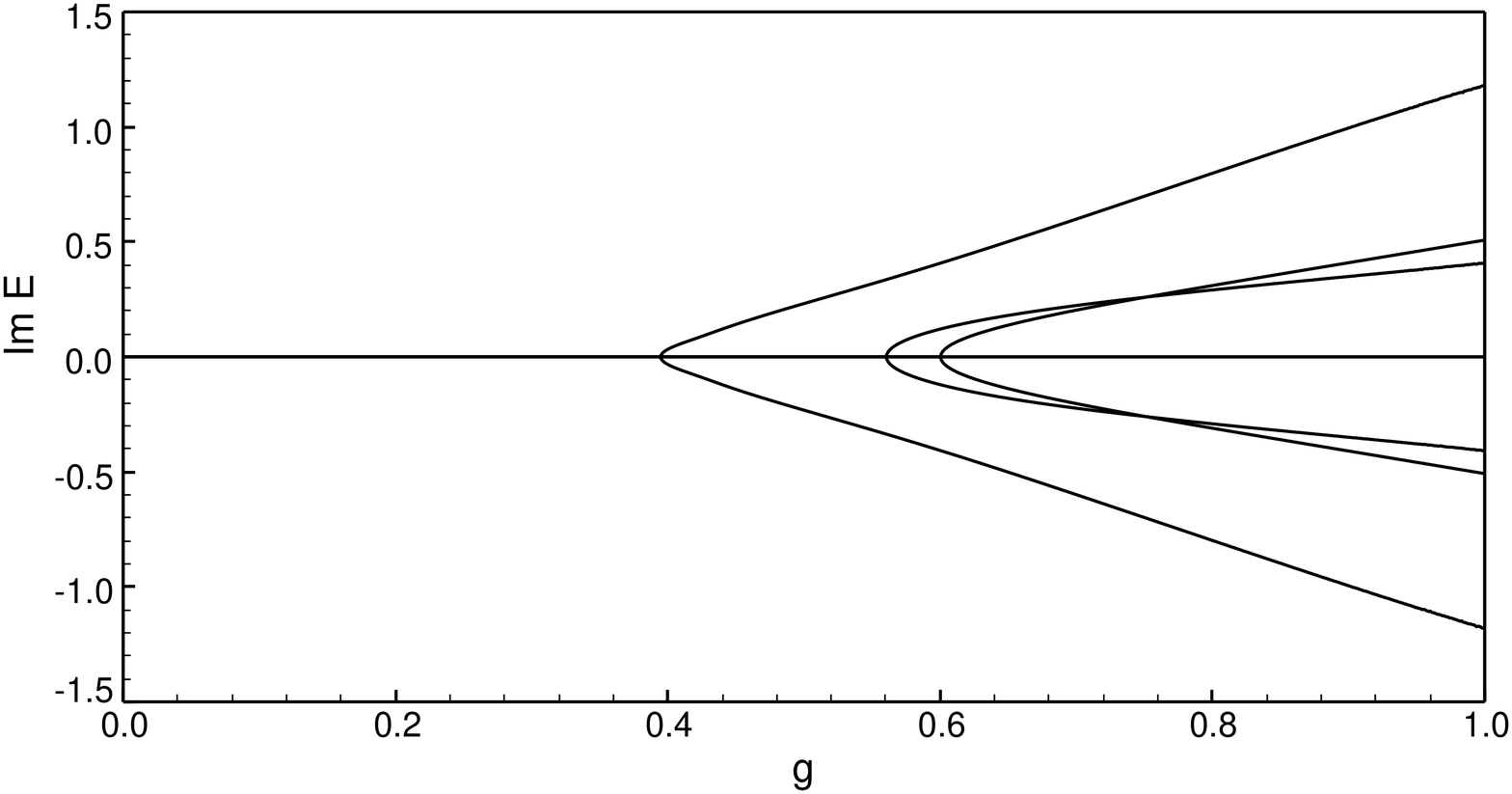}
\end{center}
\caption{Lowest eigenvalues with $m=0$ for the potential
$V(r,z)=r+igz$} \label{fig:rzm0}
\end{figure}

\begin{figure}[H]
\begin{center}
\includegraphics[width=6cm]{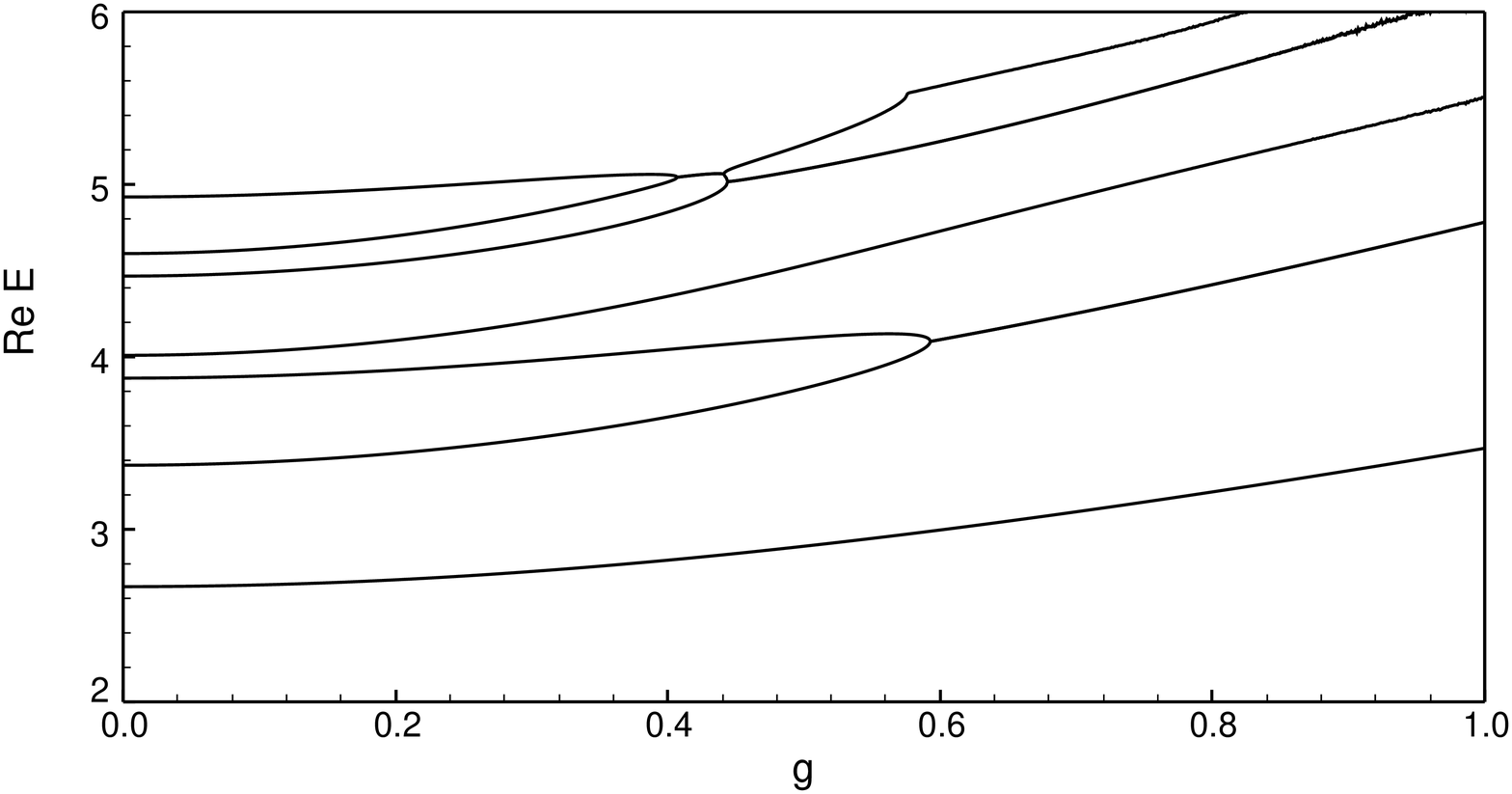} %
\includegraphics[width=6cm]{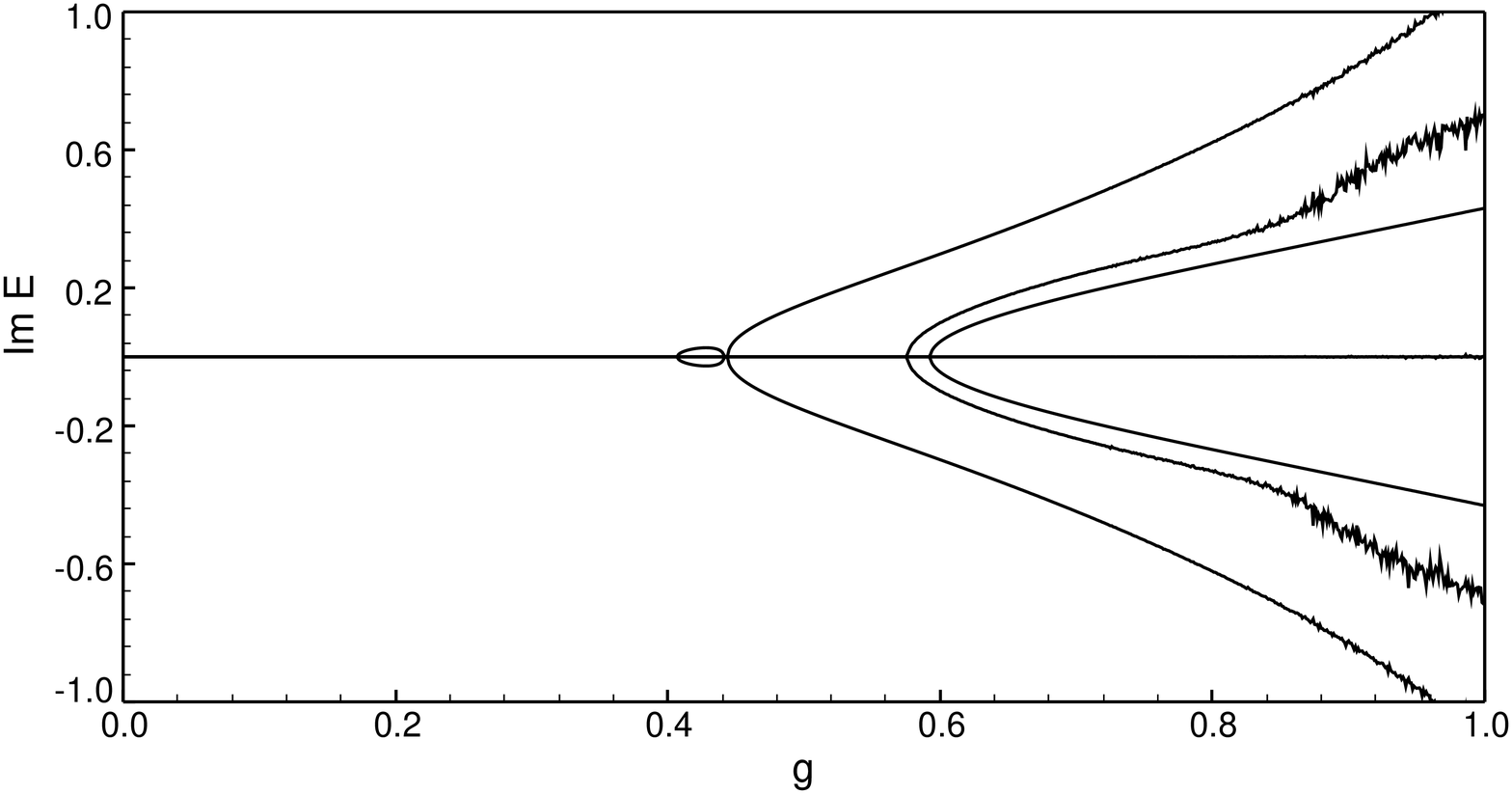}
\end{center}
\caption{Lowest eigenvalues with $m=1$ for the potential
$V(r,z)=r+igz$ with $m=1$} \label{fig:rzm1}
\end{figure}

\begin{figure}[H]
\begin{center}
\includegraphics[width=6cm]{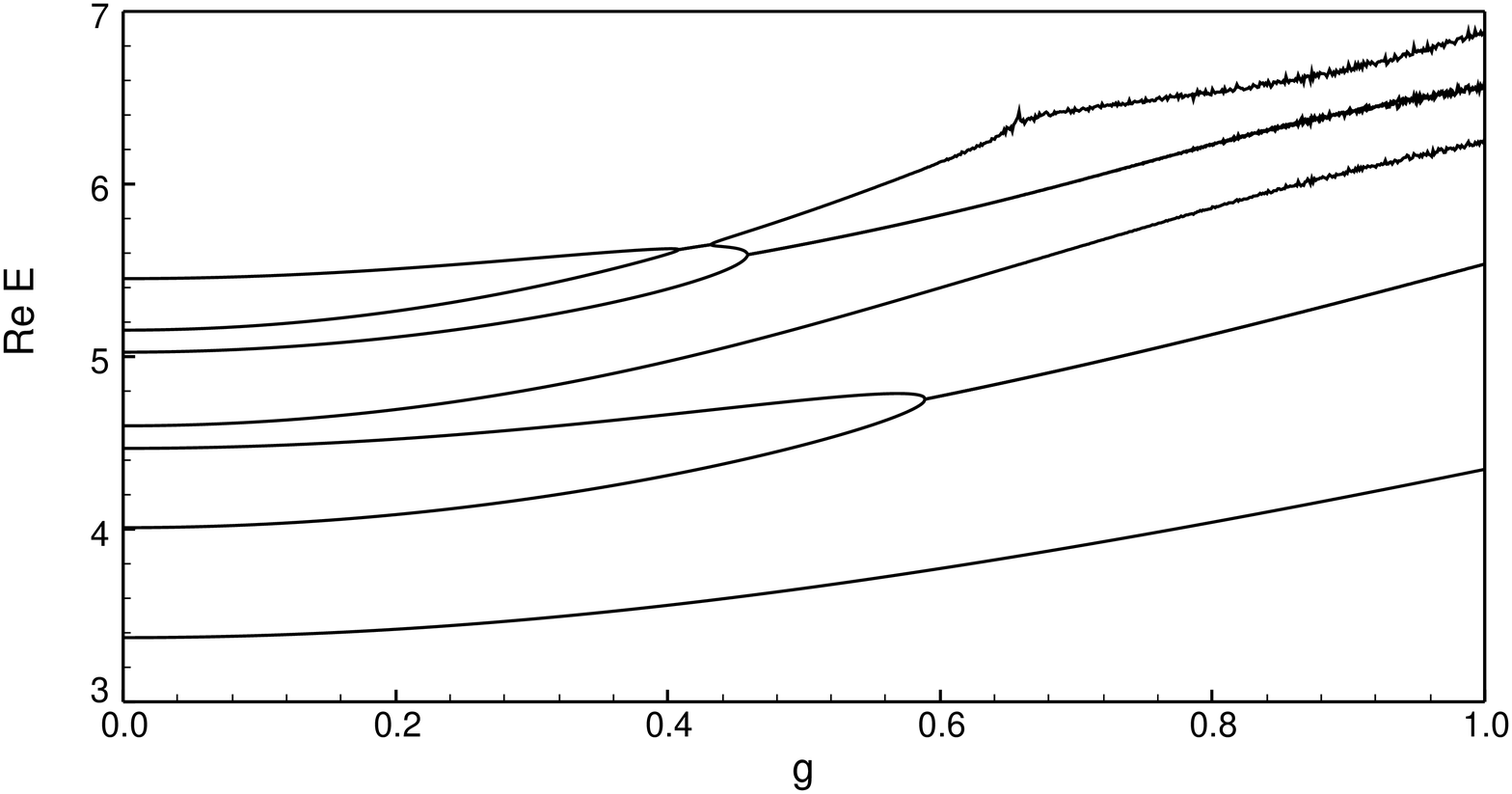} %
\includegraphics[width=6cm]{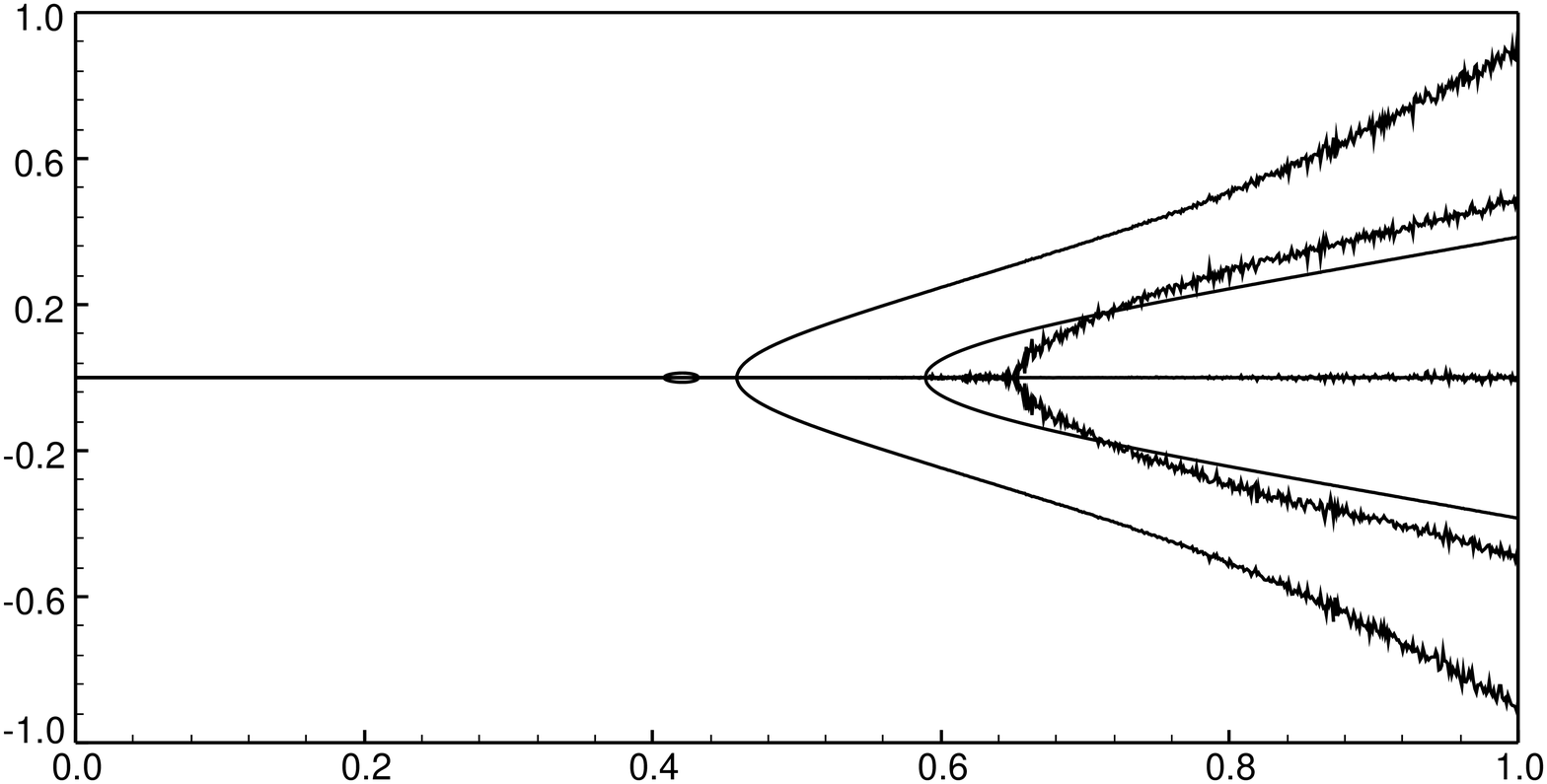}
\end{center}
\caption{Lowest eigenvalues with $m=2$ for the potential
$V(r,z)=r+igz$} \label{fig:rzm2}
\end{figure}

We can calculate approximate eigenvalues by means of diagonalization of a
suitable matrix representation of the Hamiltonian. For simplicity, here we
choose the nonorthogonal Slater-type basis set
\begin{equation}
B=\left\{ r^{n}e^{-\alpha r}Y_{l}^{m}(\theta ,\phi ),\;n,l,|m|=0,1,\ldots
\right\} .  \label{eq:basis_Slater}
\end{equation}
Present numerical results show that this problem exhibits the
usual spectral pattern common to most PT-symmetric Hamiltonians
studied by other authors; that is to say, unbroken PT symmetry for
$0<g<g_{c}$. For sufficiently small values of $g$ the eigenvalues
are real. As $g$ increases two eigenvalues approach each other,
coalesce at an exceptional point\cite{HS90,H00,HH01,H04}
$g_{i}\geq g_{c}$ becoming a pair of complex conjugate numbers for
$g>g_{i}$. This behaviour is illustrated by figures
\ref{fig:rzm0}, \ref{fig:rzm1} and \ref{fig:rzm2}, for $m=0,1,2$,
respectively. Those results were obtained by diagonalization of
the matrix representation of the Hamiltonian operator in the
Slater basis set (\ref{eq:basis_Slater}) with $\alpha =2$. The
irregular lines reflect errors in the calculation of the
eigenvalues originated in the quasi linear dependence of the basis
set. This shortcoming of the present approach becomes more
noticeable as the number of radial basis functions increases.
Although our numerical results are not extremely accurate and are
restricted to the lowest eigenvalues for the reason just
indicated, they appear to suggest that the smallest exceptional
point $g_{c}$ may be nonzero and that there is a PT phase
transition at such point. We think that a more accurate
calculation is not necessary to illustrate the difference between
this model and the other two ones discussed above.

\section{Conclusions}

\label{sec:conclusions}

In this paper we have discussed three Hamiltonians given by three
different central-field Hermitian parts and the same non-Hermitian
PT-symmetric perturbation. Although at first sight they appear to
be similar, they exhibit completely different spectra. In the case
of the isotropic harmonic oscillator the PT symmetry is unbroken
and the spectrum is real for all $g$. The reason is that $H$ and
$H_{0}$ are related by the similarity transformation
(\ref{eq:similarity}). On the other hand, the PT symmetry is
broken for all $g$ in the case of the hydrogen atom. Quite in
between the linear radial potential appears to exhibit unbroken PT
symmetry for all $0<g<g_{c}$ and a phase transition at some
$g_{c}$ that we were unable to determine.

The remarkable difference among the spectra of such seemingly similar
Hamiltonians is due to the symmetry of $H_{0}$. As a general rule the higher
the symmetry of $H_{0}$ the more likely the occurrence of complex
eigenvalues and the Hamiltonian for the hydrogen atom exhibits the greatest
symmetry by far. We have already discussed the effect of symmetry in earlier
papers\cite{FG14a,FG14b,AFG14b,AFG14c} but we have not seen such a
remarkable difference in the behaviour of the non-Hermitian Hamiltonians.

In closing we want to stress the fact that perturbation theory
provides a useful hint about the nature of the spectra of a given
non-Hermitian Hamiltonian. If a perturbation correction of odd
order (we typically look for the first one) is nonzero then we
know that the spectrum is complex for all values of $g$ (or at
least for sufficiently small $g$). If all the available
perturbation corrections of odd order are zero then there is a
chance of finding real spectrum for some values of $g$. Obviously,
this case should be investigated by more accurate calculations. As
the symmetry of $H_{0}$ increases, then also increases the
dimension of its eigenspaces and, consequently, the dimension of
the matrix representation of the perturbation in those
eigenspaces. As a result it also increases the chance of nonzero
perturbation corrections of first order.

\end{document}